\documentclass[prb,twocolumn,aps,showpacs,superscriptaddress]{revtex4}
\usepackage{amssymb, amsmath}
\usepackage{amsfonts}
\usepackage{graphicx} 
\usepackage{dcolumn} 
\usepackage{bm} 
\usepackage{textcomp}
\usepackage{appendix}

\begin{document}

\title{Surface and Edge States in Topological Semi-metals}
\author{Rui-Lin Chu, Wen-Yu Shan, Jie Lu, Shun-Qing Shen}
\affiliation{Department of Physics and Center of Computational and Theoretical Physics,
The University of Hong Kong, Pokfulam Road, Hong Kong}
\date{\today }

\begin{abstract}
We study the topologically non-trivial semi-metals by means of the 6-band
Kane model. Existence of surface states is explicitly demonstrated by
calculating the LDOS on the material surface. In the strain free condition,
surface states are divided into two parts in the energy spectrum, one part
is in the direct gap, the other part including the crossing point of surface
state Dirac cone is submerged in the valence band. We also show how
uni-axial strain induces an insulating band gap and raises the crossing
point from the valence band into the band gap, making the system a true
topological insulator. We predict existence of helical edge states and spin
Hall effect in the thin film topological semi-metals, which could be tested
with future experiment. Disorder is found to significantly enhance
the spin Hall effect in the valence band of the thin films.
\end{abstract}

\maketitle

\section{Introduction}

Topological insulators (TIs) have been recognized as novel states of quantum
matter.\cite{Moore} They are of tremendous interest to both fundamental
condensed matter physics and potential applications in spintronics as well
as quantum computing. TIs are insulating in the bulk which is usually due to
direct bulk band gaps. However, on the boundary there are topologically
protected gapless surface or edge states. Usually in these materials the
spin orbital coupling is very strong such that the conduction band and
valence band are inverted.
Such phenomena have already been reported in the literature of the 1980's.\cite{Volkov, Volkov87, Fradkin, Korenman, Agassi} However, the topological nature behind these phenomena was not revealed at the time.
With the recent development of topological band theories, a series of materials including HgTe quantum
well, Bi$_{x}$Sb$_{1-x}$, Bi$_{2}$Se$_{3}$, and Bi$_{2}$Te$_{3}$ have been
theoretically predicted and experimentally realized as 2D and 3D TIs.\cite{Kane, Qi-prb,Bernevig, Fu3D,
Fuprb,konig-science,Hsieh-nature,Haijunzhang-nat,Xia-nat-phy} The search for
TIs has been extended from these alloys and binary compounds to ternary
compounds. Very recently a large family of materials namely the Heusler-related
and Li-based intermetallic ternary compounds have been predicted to be
promising 3D TIs through first principle calculations.\cite{XiaoDi-prl,Franz-view, Chadov,HLin,HLin2,HLin3, BHYan} The enormous variety
in these compounds provides wide options for future material synthesizing of
TIs. However, despite an insulating bulk is a critical prerequisite to the
TI theory, none of these newly found materials are naturally band
insulators. Many of them are semi-metals or even metals. It has been shown
that the band structure and band topology of many of these compounds closely
resemble that of the zinc blende structure binary compound HgTe and CdTe.
\cite{XiaoDi-prl,Franz-view,Chadov,HLin,HLin2,HLin3,BHYan}

In this work we study the evolution of surface states in these topologically
non-trivial compounds whose band structure closely resemble 3D HgTe by means
of the 6-band Kane model. By studying the local density of states (LDOS) on
the material boundary, we demonstrate explicitly the existence of surface
states in the direct band gap. Furthermore, we show that in the strain free
condition, the surface states are separated into two parts in the band
structure. One part exists in the direct band gap, the other part of surface
states (including the crossing point of Dirac cone) submerge in the valence
bands. The latter surface states have distinct momentum dependent spatial
distribution from the former. By applying uniaxial strains, the crossing
point can be raised up into the strain induced insulating gap. In the thin
films made of these materials, topologically protected helical edge states
will emerge on the sample edges. In this way we show a crossover between 3D
TIs and 2D TIs. In the strain free condition, the bulk band gap can be
controlled by tuning the film thickness. Meanwhile, we found
that in the thin films disorders or impurities can significantly enhance the
spin Hall effect(SHE) when the Fermi level is in the valence band.

\begin{figure}[tbph]
\centering \includegraphics[height=45mm]{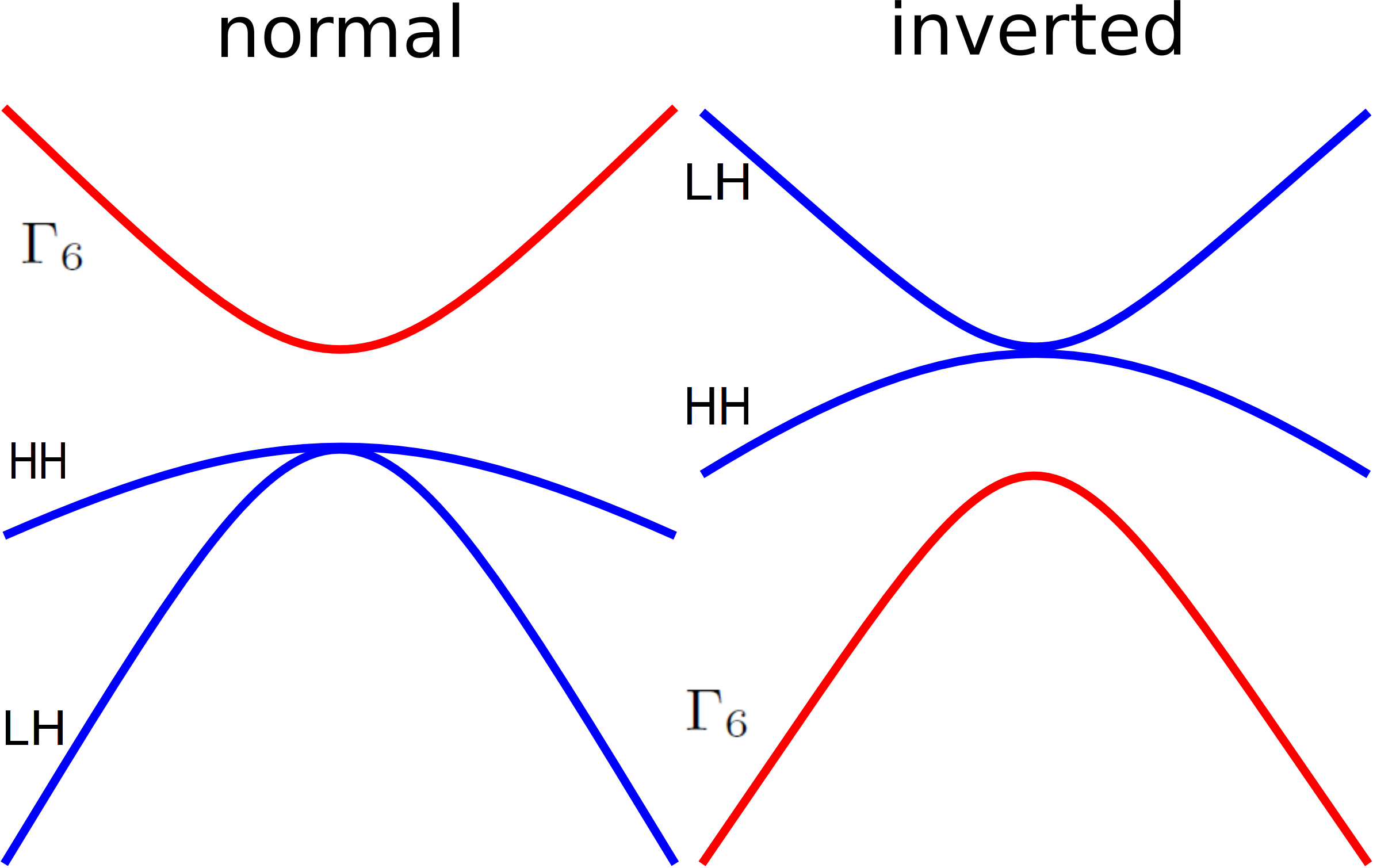}
\caption{(color online) A schematic illustration of the band inversion between $\Gamma_6$
(red curve) and $\Gamma_8$(blue curve), the left is the normal case where
the blue curve represents the LH and HH of $\Gamma_8$ valence band, the
right is the inverted case where the LH flips up and becomes the conduction
band, the $\Gamma_6$ appears below the HH band.}
\end{figure}

\section{Model Hamiltonian}

3D HgTe and CdTe share the same zinc blende structure. The band topology of
these two materials is distinguished by the band inversion at the $\Gamma $
-point, which happens in HgTe but not CdTe. This causes HgTe to be
topologically nontrivial while CdTe is trivial.\cite{Volkov87, Fuprb, Daixiprb} The
essential electronic properties of both are solely determined by the band
structure near the Fermi surface at $\Gamma $-point, where the bands possess 
$\Gamma _{6}$ ($s$ type, doubly degenerate), $\Gamma _{8}$ ($p$ type, $j=3/2$
, quadruply degenerate) and $\Gamma _{7}$ ($p$ type, $j=1/2$, doubly
degenerate) symmetry.\cite{Novik} The band inversion in HgTe takes place
because the $\Gamma _{6}$ bands appear below the $\Gamma _{8}$ band, whereas
in the normal case (such as CdTe) $\Gamma _{6}$ is above $\Gamma _{8}$ (see
Fig 1).\cite{Winkler,kane-model} In this work, we use a 6-band Kane model
Hamiltonian which takes into account the $\Gamma _{6}$ and $\Gamma _{8}$
band. The spin-orbit split off $\Gamma _{7}$ band usually appears far below
the $\Gamma _{6}$ and $\Gamma _{8}$ bands and hence can be neglected because
it doesn't affect the low energy approximation. The 6-band Kane model
describes the band structure near the $\Gamma $-point well and adequately
captures the band inversion story.\cite{Volkov87,Daixiprb} As expected, we find
gapless surface states exist in the direct gap of 3D HgTe while absent in
CdTe.

\begin{figure*}[htp]
\centering \includegraphics[height=50mm]{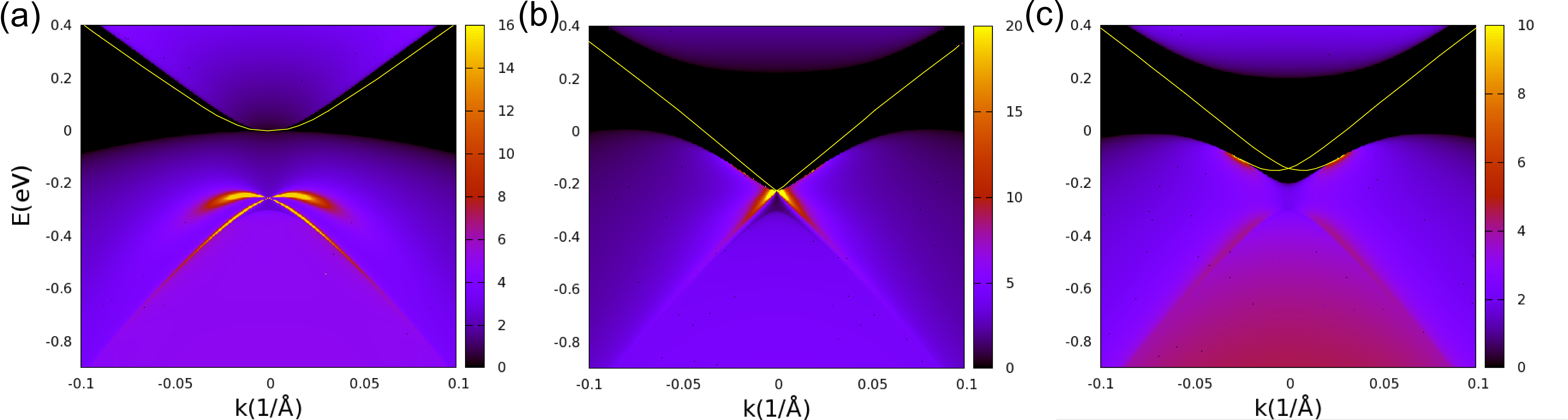}
\caption{(color online) (a) Surface LDOS of 3D HgTe without strain, bright line in the
direct gap between LH and HH $\Gamma_8$ bands indicates the first type
surface state, bright regions in the valence band indicates the second type
surface state, $E_{\Gamma_6} < E_0 < E_{HH} = E_{LH}$; (b) A
insulating band gap is opened with strain $T_{\protect\epsilon}=0, U_{
\protect\epsilon}=0, V_{\protect\epsilon}=-0.224 eV$, the first and second
type surface states become connected, $E_{\Gamma_6}<E_{HH}=E_0<E_{LH}$; (c) same with (b) at $T_{\protect
\epsilon}=0.1 eV, U_{\protect\epsilon}=-0.05 eV, V_{\protect\epsilon}=-0.25
eV$, note that the $\Gamma_6$ band has been inverted with the HH $\Gamma_8$
band and thus appears in the middle of $\Gamma_8$ bands(LH and HH), i.e. $E_{HH}<E_{\Gamma_6}<E_0<E_{LH}$. A highest LDOS
limit is set and higher data points have been filtered to this limit for a
clearer view of the whole spectrum.}
\end{figure*}

The 6-band Kane Hamiltonian writes: 
\begin{widetext}
\begin{eqnarray}\label{Hamiltonian1}
H_0=\left(\begin{array}{cccccc}
T & 0  &  -\frac{1}{\sqrt{2}}Pk_+  & \sqrt{\frac{2}{3}}Pk_z &  \frac{1}{\sqrt{6}}Pk_- & 0   \\
\\

0 & T & 0  &  -\frac{1}{\sqrt{6}}Pk_+  & \sqrt{\frac{2}{3}}Pk_z & \frac{1}{\sqrt{2}}Pk_- \\
\\

-\frac{1}{\sqrt{2}}Pk_- & 0 & U+V & 2\sqrt{3}\bar{\gamma} Bk_-k_z  &  \sqrt{3}\bar{\gamma}Bk_-^2 & 0 \\
\\

\sqrt{\frac{2}{3}}Pk_z & -\frac{1}{\sqrt{6}}Pk_- & 2\sqrt{3}\bar{\gamma} Bk_+k_z & U-V & 0 & \sqrt{3}\bar{\gamma}Bk_-^2   \\
\\
\frac{1}{\sqrt{6}}Pk_+ & \sqrt{\frac{2}{3}}Pk_z & \sqrt{3}\bar{\gamma}Bk_+^2 &  0 & U-V & -2\sqrt{3}\bar{\gamma}Bk_-k_z  \\
\\
0 & \frac{1}{\sqrt{2}}Pk_+ & 0 & \sqrt{3}\bar{\gamma} Bk_+^2 & -2\sqrt{3}\bar{\gamma}Bk_+k_z & U+V   \\
\end{array}\right),
\end{eqnarray}
\end{widetext}where 
\begin{eqnarray}
&&B=\frac{\hbar^2}{2m_0},  \notag \\
&&P=\sqrt{\frac{\hbar^2E_p}{2m_0}},  \notag \\
&&k_{\Vert }^{2}=k_{x}^{2}+k_{y}^{2},\ \ k_{\pm }=k_{x}\pm ik_{y},  \notag \\
&&T=E_{g}+B(2F+1)(k_{\Vert }^{2}+k_{z}^{2}),  \notag \\
&&U=-B\gamma _{1}(k_{\Vert }^{2}+k_{z}^{2}),  \notag \\
&&V=-B\bar{\gamma}(k_{\Vert }^{2}-2k_{z}^{2}),  \notag
\end{eqnarray}
where $m_0$ is the electron mass. For the simplicity of the physical picture, we have taken the axial
approximation $\bar{\gamma}=(\gamma _{2}+\gamma _{3})/2$ which makes the
band structure isotropic in $k_{x},k_{y}$ plane. $E_{g}$, $E_{p}$, $F$, $
\gamma _{1}$, $\gamma _{2}$ and $\gamma _{3}$ are material specific
parameters. The basis functions are denoted as $(|\psi _{1}\rangle ,|\psi
_{2}\rangle )\Gamma _{6},(|\psi _{3}\rangle ,|\psi _{4}\rangle ,|\psi
_{5}\rangle ,|\psi _{6}\rangle )\Gamma _{8}$. Here we take the parameters of
HgTe just for illustration of the physics (Table.1). The same physics should
also happen in the recently discovered Heusler and Li-based ternary
compounds which share similar band topologies\cite
{XiaoDi-prl,Franz-view,Chadov,HLin,HLin2,HLin3,BHYan}. 
\begin{table}[tbph]
\caption{ Band structure parameters of HgTe at $T=0$K.\protect\cite{Novik}}
\label{tab:HgTeparameters}
\begin{ruledtabular}
\begin{tabular}{cccccccc}
   $E_g$ & $E_P=2m_0P^2/\hbar^2$ & $F$ & $\gamma_1$  & $\gamma_2$ & $\gamma_3$   \\
  \hline
\\
   -0.3 eV     & 18.8 eV   & 0 & 4.1  & 0.5  & 1.3     \\
\end{tabular}
\end{ruledtabular}
\end{table}

\section{Two types of surface states in 3D semi-metals}

Without strain, HgTe is a semi-metal with an inverted band structure. At the $
\Gamma $ point, the conduction band ($\Gamma _{8}$ LH) touches the valence
band ($\Gamma _{8}$ HH) and the $\Gamma _{6}$ is below the $\Gamma _{8}$
band (as in Fig.1).
In Hamiltonian $H_0$ the band topology is solely determined by the parameter $E_{g}$. At the $\Gamma$ point,
LH and HH are degenerate at energy $E=0$, and $\Gamma_6$ is at position $E=E_g$. So when $
E_{g}>0$ the band structure is normal and insulating with a positive energy gap $\Delta=E_g$. When $E_{g}<0$ the
band structure is inverted $\Gamma_6$ appears below LH and HH. In this 
case, the system is semi-metal since the bulk band gap $\Delta$ is always 0 because of the LH/HH degeneracy. We will show this semi-metal is topologically non-trivial by that gapless surface states exists in the direct gap of LH and HH. And by lifting the degeneracy of LH and HH by strain or finite confinement which makes $\Delta$ non-zero, the system instantly becomes a topological insulator. Whereas in trivial semi-metals, lifting the zero band gap only turns the systems into trivial band insulators.

To study the topological properties of this Hamiltonian we transform it into
a tight-binding model on a cubic lattice, where such approximation
substitutions are used: 
\begin{eqnarray}
&&k_{i}\rightarrow \frac{1}{a}sin(k_{i}a),  \notag  \label{tbsub} \\
&&k_{i}^{2}\rightarrow \frac{2}{a^{2}}(1-cos(k_{i}a)),
\end{eqnarray}
here $k_{i}$ refers to $k_{x},k_{y}$ and $k_{z}$, a is the lattice constant
which is taken to be 4 \AA {} in this work. This approximation
is valid in the vicinity of $\Gamma$ point.

Surface states reside only on the system surface, which will project larger
LDOS on the surface than the bulk states. Thus the surface LDOS can be
studied to identify the existence of surface states. The surface LDOS is
given by $\rho (k)=-\frac{1}{\pi }$Tr$ImG_{00}(k)$, where $G_{00}$ is the
retarded Green's function for the top layer of a 3D lattice. In a
semi-infinite 3D system the surface Green's function can be obtained by means of the transfer matrix,
\begin{eqnarray}
G_{00}&=&(E-H_{00}-H_{01}\mathbf{T})^{-1}, \label{green1} \\
\mathbf{T}&=&(E-H_{00}-H_{01}\mathbf{T})^{-1}H_{01}^{\dag},\label{green2}
\end{eqnarray}
where $H_{00}$ and $H_{01}$ are the Hamiltonian elements within and between the layers(or super cells) and $\mathbf{T}$ is the transfer matrix. Usually Eq(\ref{green2}) can be calculated iteratively until $\mathbf{T}$ converges, which is quite time consuming. Here we use a fast converging algorithm proposed by Sancho et,el to calculate the transfer matrix.\cite{Sancho} In Fig.2a we present the
LDOS on an infinite xy surface, where z dimension is semi-infinite.  As expected, the LDOS clearly shows the existence of
surface states between the LH and HH bands. Through further checking we
find that these states indeed reside only on the surface boundary (see Fig.
3a). Its spatial distribution is in the decaying form beneath the surface.
Interestingly, we also find another kind of surface state submerging in the
valence bands (bright regions in the valence band of Fig. 3a). It shows up
between the inverted $\Gamma _{6}$ and HH $\Gamma _{8}$ bands. We confirm
its surface state nature by checking its spatial distribution along the z
direction (Fig. 3b and 3c). It is found that the spatial distribution of
this surface state has very distinct form from that between the LH and HH bands. It
bears the oscillating feature of LDOS of bulk states (see Fig. 3d) but is
also clearly decaying beneath the surface. We call the former surface state
as the first type and the latter the second type. Both types of surface
state project much larger LDOS on the surface than the bulk states. However,
the first type surface state is clearly decoupled from the bulk states while
the second type is coupled with the valence bulk conduction states.

The spatial distribution can be obtained by enlarging the super cell when
calculating the surface Green's function $G_{00}$. For example, take the
first 50 layers as a unit cell, we can obtain the LDOS distribution in the
first 50 layers. Fig. 3a shows the spatial distribution of the first type
surface state at variant $k$. It shows that the closer to the $\Gamma $
point the wider it distributes in space. Only away from the $\Gamma$ point, it showes strong localization
on the surface. Close to the $\Gamma $
point, its distribution is bulk like and can barely be recognized as surface
state. In Fig. 2 the bright line indicating first type surface
has been highlighted for a clearer view. The Fermi surface of
3D HgTe is close to the point where the conduction band touches the valence
band.\cite{Chadov,Tsidi} The first type surface states at the Fermi surface
distribute widely in the space. They are expected to make no significant
contribution to the transport properties either. Without appropriate doping
or gating it is difficult to detect the first type surface states in 3D HgTe
experimentally either through angle-resolved photo-emission spectroscopy
(ARPES) or transport measurements.

Unlike the first type surface states which show up as well defined sharp
lines in the energy spectrum, the second type surface states appear as two
crossing bright regions submerging in the valence bands in the energy
spectrum. It appears near the boundary of the $\Gamma _{6}$ band. At the
crossing point, it is the sharpest, above the crossing point, it is
broadened into two wide tails. Close to the $\Gamma $ point, its LDOS on the
surface is the highest. Away from the $\Gamma $ point, the surface LDOS
becomes smaller and finally those states merge into the bulk states. Also
unlike the first type surface state whose distribution width becomes very
large when approaching the $\Gamma $ point, its distribution width in space
does not significantly depend on the momentum. At the $\Gamma$ point, the
second type surface state energy and wave function is exactly solvable from
the Hamiltonian. We will show that this point is the crossing point of the surface state Dirac cone.(see text below)
 Beneath the surface, the LDOS quickly
decays in an oscillatory way. For comparison, we also plot out the bulk
states LDOS which is oscillatory and extended all over the space without
decaying (Fig. 3d). Since from the LDOS the second type surface state can be
easily distinguished from the bulk states, we believe they may be easily
detected from the ARPES measurements as well.

\begin{figure}[t]
\centering \includegraphics[width=88mm]{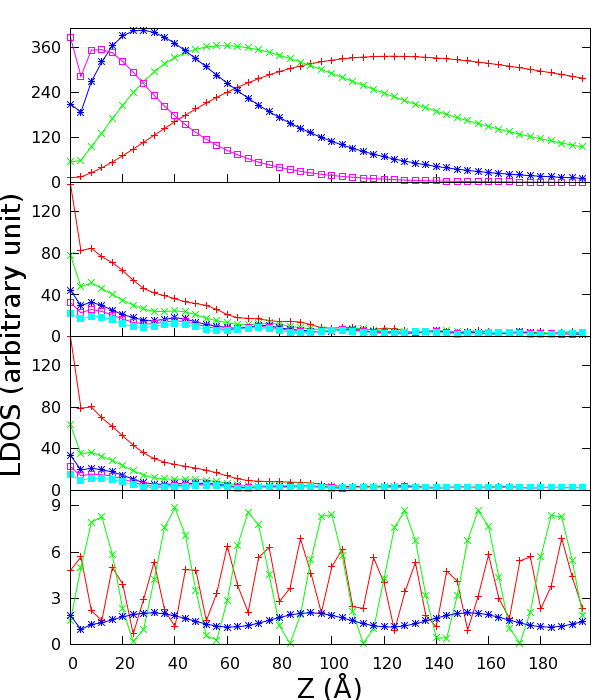}
\caption{(color online) Surface state(ss) distribution along z direction. (a)First type ss
at k=0.005(red), 0.01(green), 0.02(blue), 0.035(pink) $1/\mathring{A}{}$;
(b) second type ss below the crossing point at k=0.0066(red), 0.01(green),
0.014(blue), 0.017(pink), 0.022(cyan) $1/\mathring{A}{}$; (c) second type ss
above the crossing point at k same as in (b); (d) bulk states at k=0.0066 $1/
\mathring{A}{}$ and E=-0.4eV(red), -0.1eV(green), 0.2eV(blue).}
\end{figure}

\section{Origin of surface states and Strain induced band gap}

The origin of the two types of surface state can be understood intuitively
in the following way. If we divide the Hamiltonian into two subspaces which
are spanned by two groups of basis $\mathcal{P}_{1}=\{\psi _{1},\psi _{4},\psi
_{2},\psi _{5}\}$ and $\mathcal{P}_{2}=\{\psi _{3},\psi _{6}\}$, it is
amazing that the effective Hamiltonian in the subspace $\mathcal{P}_{1}$ is
quite similar to the effective model proposed for a 3D topological insulator.
\cite{Haijunzhang-nat} The Hamiltonian after arranging the basis as ($
\mathcal{P}_{1}(\psi _{1},\psi _{4},\psi _{2},\psi _{5}),\mathcal{P}
_{2}(\psi _{3},\psi _{6}$)) reads 
\begin{widetext}
\begin{eqnarray}\label{Hamiltonian2}
H_0'=\left(\begin{array}{cccccc}
T & \sqrt{\frac{2}{3}}Pk_z  &  0  &  \frac{1}{\sqrt{6}}Pk_- & -\frac{1}{\sqrt{2}}Pk_+ & 0   \\
\\
\sqrt{\frac{2}{3}}Pk_z & U-V & -\frac{1}{\sqrt{6}}Pk_- & 0&2\sqrt{3}\bar{\gamma} Bk_+k_z &\sqrt{3}\bar{\gamma}Bk_-^2   \\
\\
0 &  -\frac{1}{\sqrt{6}}Pk_+  & T & \sqrt{\frac{2}{3}}Pk_z & 0& \frac{1}{\sqrt{2}}Pk_- \\
\\
\frac{1}{\sqrt{6}}Pk_+ & 0& \sqrt{\frac{2}{3}}Pk_z & U-V &\sqrt{3}\bar{\gamma}Bk_+^2 & -2\sqrt{3}\bar{\gamma}Bk_-k_z  \\
\\

-\frac{1}{\sqrt{2}}Pk_-  & 2\sqrt{3}\bar{\gamma} Bk_-k_z  & 0&   \sqrt{3}\bar{\gamma}Bk_-^2 & U+V &  0 \\
\\

0 &  \sqrt{3}\bar{\gamma} Bk_+^2 & \frac{1}{\sqrt{2}}Pk_+ & -2\sqrt{3}\bar{\gamma}Bk_+k_z & 0& U+V   \\
\end{array}\right),
\end{eqnarray}
\end{widetext}Without considering the coupling between $\mathcal{P}_{1}$ and 
$\mathcal{P}_{2}$ sub-spaces, $\mathcal{P}_{1}$ gives the LH (conduction)
and $\Gamma _{6}$ (valence) band, plus a gapless Dirac cone of surface states
between their gap; $\mathcal{P}_{2}$ gives the HH (valence) which overlaps
with the surface state Dirac cone and $\Gamma _{6}$ band. After turning on
the coupling between the two sub-spaces, the surface state Dirac cone
becomes separated into two parts. One part appears in the direct gap between
LH and HH, the other submerges into the HH of valence band. In this sense, our results are very similar 
to an earlier work which discusses interface states in $Hg_{1-x}Cd_xTe$ heterojunctions.\cite{Volkov87}

The $\Gamma_8$ degeneracy at the $\Gamma$ point makes 3D HgTe a
semi-metal. To lift the degeneracy and open an insulating gap at the Fermi
energy, which makes the system a true 3D TI, we consider applying a uniaxial
strain along the $(001)$ axis.\cite{Fuprb,Daixiprb} According to Ref.\cite
{Winkler}, the additional strain induced Hamiltonian is introduced as (in
the original basis order as in Eq. (1)) 
\begin{equation*}
H_{s}=\left( 
\begin{array}{cccccc}
T_{\epsilon } & 0 & 0 & 0 & 0 & 0 \\ 
0 & T_{\epsilon } & 0 & 0 & 0 & 0 \\ 
0 & 0 & U_{\epsilon }+V_{\epsilon } & 0 & 0 & 0 \\ 
0 & 0 & 0 & U_{\epsilon }-V_{\epsilon } & 0 & 0 \\ 
0 & 0 & 0 & 0 & U_{\epsilon }-V_{\epsilon } & 0 \\ 
0 & 0 & 0 & 0 & 0 & U_{\epsilon }+V_{\epsilon }
\end{array}
\right) ,
\end{equation*}
where the $T_{\epsilon }$, $U_{\epsilon }$ and $V_{\epsilon }$ are strain
induced interaction terms. To illustrate the physical picture more clearly,
we have made these terms artificially large in this work. At the $\Gamma$
point, the boundary for the $\Gamma _{6}$ band is $E_{\Gamma_6}=E_{g}+T_{\epsilon }$, for
HH band is $E_{HH}=U_{\epsilon }+V_{\epsilon }$ and for LH band is $E_{LH}=U_{\epsilon
}-V_{\epsilon }$. In the strain free condition, $E_{\Gamma_6} < E_{LH} = E_{HH} = 0$. 
However, under strain, the relative
position of the three is subject to change and is determined by the strain terms.

Surface state at the crossing point of the Dirac cone is easily solvable
directly from the Hamiltonian $H_{0}+H_{s}$, where $k_{x}=k_{y}=0$.\cite
{shanwy} Using the boundary condition $\psi (z=0)=0$ and $\psi (z=-\infty
)=0 $, we get the eigen energy for surface state 
\begin{eqnarray}
E_{0} &=&C+\frac{D_{1}M}{B_{1}}  \notag \\
&=&\frac{(E_{g}+T_{\epsilon })(\gamma _{1}+2\overline{\gamma })+(U_{\epsilon
}-V_{\epsilon })(2F+1)}{\gamma _{1}+2\overline{\gamma }+2F+1}
\end{eqnarray}
where 
\begin{eqnarray}
C &=&\frac{E_{g}+T_{\epsilon }+U_{\epsilon }-V_{\epsilon }}{2},  \notag \\
M &=&\frac{E_{g}+T_{\epsilon }-U_{\epsilon }+V_{\epsilon }}{2}.  \notag \\
D_{1} &=&\frac{B}{2}(2F+1-\gamma _{1}-2\overline{\gamma }),  \notag \\
B_{1} &=&-\frac{B}{2}(2F+1+\gamma _{1}+2\overline{\gamma }).
\end{eqnarray}
And the wave function is 
\begin{equation*}
\varphi =\frac{1}{\sqrt{2}}\left( 
\begin{array}{c}
i\sqrt{\frac{D_{+}}{B_{1}}} \\ 
\sqrt{-\frac{D_{-}}{B_{1}}}
\end{array}
\right) c(e^{\lambda _{1}z}-e^{\lambda _{2}z}),
\end{equation*}
with the basis $\psi _{1}$ and $\psi _{4}$ or $\psi _{2}$ and $\psi _{5}$,
where 
\begin{eqnarray}
\lambda _{1,2}^{2} &=&\frac{D_{2}\pm \sqrt{D_{2}^{2}-D_{3}}}{2D_{+}D_{-}}
,~with  \notag \\
D_{2} &=&-[A_{1}^{2}+D_{+}(E_{0}-L_{1})+D_{-}(E_{0}-L_{2})],  \notag \\
D_{3} &=&4D_{+}D_{-}(E_{0}-L_{1})(E_{0}-L_{2}),  \notag
\end{eqnarray}
$D_{\pm }=D_{1}\pm B_{1}$, $L_{1}=C+M$, $L_{2}=C-M$ and $A_{1}=\sqrt{\frac{2
}{3}}P$. For the surface state solution to exist, it is required that: 
\begin{equation*}
MB_{1}>0.
\end{equation*}
This condition is easily satisfied in an inverted band structure without
strain, where $E_{g}<0$.

In Fig. 2b and 2c the spectrum after adding the strain interaction is
plotted. An insulating gap is opened between the LH and HH $\Gamma _{8}$
bands. Meanwhile, both the first and second type surface state go through an
evolution. At a critical point (Fig. 2b) the crossing point can move to the
top to the valence band at the $\Gamma $ point and both types of surface
states become connected with each other. On further increasing the strain
strength, the crossing point jumps out of the valence band and sits in the
band gap. At this stage, we find only one kind of surface state appearing in
the insulating gap (Fig. 2c). The spectrum of surface state forms a gapless
Dirac cone at the $\Gamma $ point and the system becomes a typical 3D TI. Notice that Fig.2c closely resembles 
the band structure of 3D TI $Bi_2Se3$ and $Bi_2Te_3$.\cite{Haijunzhang-nat,Xia-nat-phy}
Although Fig. 2 is obtained from the tight binding model on a lattice, we
find the energy of the crossing point agrees with the exact solution quite
precisely.

\section{Edge states in 2D thin films}

\subsection{Quasi-2D lattice model}

It is however technically difficult applying strong enough strains to make a
semi-metal insulating in experimental conditions. The finite size effect can
open a gap in the bulk states at the $\Gamma $ point but will also open a
gap in the surface states.\cite{Zhouprl,shanwy} Instead of considering the
surface states, we make the 3D topologically non-trivial semi-metals into
thin films and study their edge effects in the strain free condition. The
same lattice model is used as in the 3D case, but now z dimension is of
finite thickness $L$. We study the LDOS on the side surface of a thin film
where x is infinite and y is semi-infinite. In this case, only $k_{x}$ is
good quantum number. In Fig.4 the LDOS on the film edge surface for various
film thickness is plotted. In the thick film limit, the band gap opened by
finite size effect is not obvious. The spectrum still resembles that of 3D
case(Fig.4a and 4b). Surface states still show residuals on the spectrum.
The finite confinement of z causes energy level discretization as in a
quantum well, which induces a series of sub-bands appearing in the energy spectrum as layered structures.
When the film is thinned down the discretized energy level spacing
increases. A pair of edge states with linear dispersion are found in the
band gap(Fig.4c/d). The system becomes a 2D topological insulator. When the
film is thinner than $20\mathring{A}{}$, another transition happens, the
system becomes trivial. The edge states disappear from the band gap.

\begin{figure}[t]
\centering \includegraphics[width=90mm]{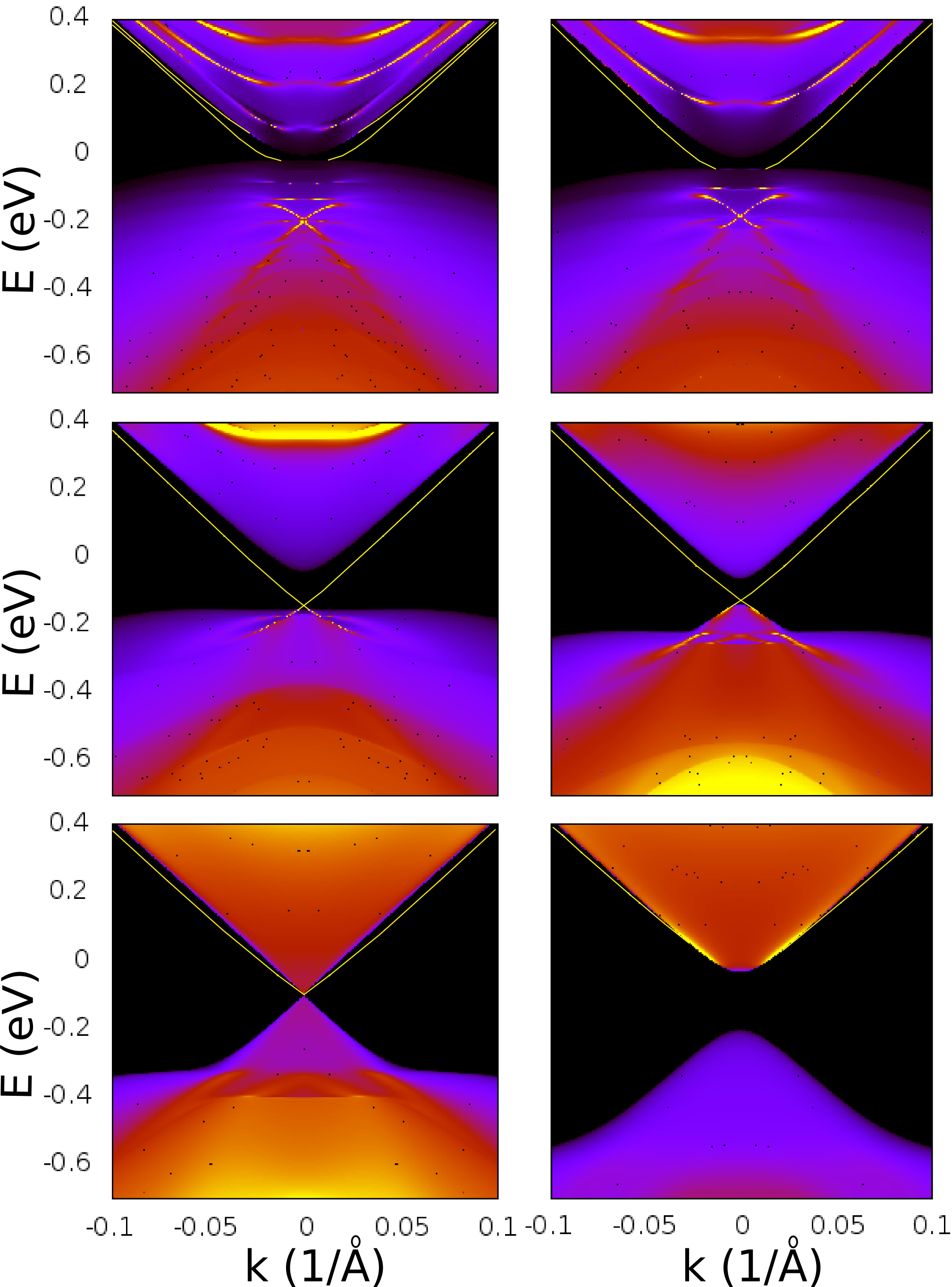}
\caption{(color online) LDOS at the edge of thin films at different thickness calculated
with 3D lattice model. (a) L=116 \AA {}; (b) L=76 \AA {}; (c) L=36 \AA {};
(d) L=28 \AA {}; (e) L=20 \AA {}; (f) L=12 \AA {}. }
\end{figure}

\subsection{Quantum well approximation and 2D lattice model}

When the film is thin enough, the finite size caused band gap becomes obvious. In this case the finite confinement  induced sub-bands are far away from the low energy regime. We can then use the quantum well approximation $\langle k_z \rangle =0,~\langle k_z^2 \rangle \simeq (\pi/L)^2$.\cite{Bernevigprl} Using these relations in the Hamiltonian in Eq. (\ref{Hamiltonian1}), and choosing the basis set in the sequence ($|\psi_1\rangle$, 
$|\psi_3\rangle$, $|\psi_5\rangle$, $|\psi_2\rangle$, $|\psi_6\rangle$, $|\psi_4\rangle$), we can obtain a two dimensional 6-band Kane model
\begin{eqnarray} \label{2Dmodel}
H(\mathbf{k})=\left(\begin{array}{cc}
                   h(\mathbf{k}) & 0 \\
                   0       & h^*(-\mathbf{k}) \\
                  \end{array}\right),
\end{eqnarray}
where 
\begin{widetext}
\begin{eqnarray}\label{upperHamiltonian}
h(\mathbf{k})=\left(\begin{array}{ccc}
                   E_g+B(2F+1)(k_{\|}^2+\langle k_z^2\rangle) & -\frac{1}{\sqrt{2}}Pk_+ & \frac{1}{\sqrt{6}}Pk_-  \\
                   -\frac{1}{\sqrt{2}}Pk_- & -(\gamma_1+\overline{\gamma})Bk_{\|}^2-(\gamma_1-2\overline{\gamma})B\langle k_z^2\rangle & \sqrt{3}\overline{\gamma}Bk_-^2 \\
                   \frac{1}{\sqrt{6}}Pk_+ & \sqrt{3}\overline{\gamma}Bk_+^2 & -(\gamma_1-\overline{\gamma})Bk_{\|}^2-(\gamma_1+2\overline{\gamma})B\langle k_z^2\rangle \\
                  \end{array}\right).
\end{eqnarray}
\end{widetext}The system keeps time-reversal symmetry, and the
representation of the symmetry operation in the new set of basis is given by 
$T=\mathcal{K}\cdot i\sigma ^{y}\otimes I_{3\times 3}$, where $\mathcal{K}$
is the complex conjugation operator, $\sigma ^{y}$ and $I$ denote the Pauli
matrix and unitary matrix in the spin and orbital space respectively. We can
study the two blocks separately since they are time-reversal counterparts of
each other. Here we focus on the upper block first. At $k_{x}=0$, the
boundaries of $\Gamma _{6}$,LH and HH are at $E=E_{g}+B(2F+1)\langle
k_{z}^{2}\rangle $,$E=-(\gamma _{1}-2\overline{\gamma })B\langle
k_{z}^{2}\rangle $, and $E=-(\gamma _{1}+2\overline{\gamma })B\langle
k_{z}^{2}\rangle $, which are controllable by choosing film thickness $L$. When $L$ decreases from the thick limit,
Down to $L\approx 30\mathring{A}{}$ $\Gamma _{6}$ band flips up and
exchanges position with HH, the system is still non-trivial. Further down to 
$L\approx 20\mathring{A}{}$, $\Gamma _{6}$ flips up and exchanges with the
conduction band(LH)(see Fig.5e). The band structure then becomes trivial just as the illustrated case in
Fig.1. This rough picture serves as an intuitive understanding of the topological transition and edge state formation in the thin films. It also agrees with result we obtained with 3D lattice model in Fig.4. 
In the thick film limit, the bulk band gap is always between LH and HH derived states while in thin limit it is between $\Gamma_6$ and LH derived states. In the thick film limit, surface states(on top and bottom surfaces) and edge states(on side surfaces) co-exists. The first type surface states on the top/bottom surfaces become the effective bulk states of the film appearing as conduction band minimum. The edge states in the thinner films actually evolve out from the surface states on the side surfaces, whose nature is 2D instead of 1D.

Using the same approximation in eq.(\ref{tbsub}), we
can transform $h(\mathbf{k})$ into a tight binding model on a 2D lattice. In
Fig.5 we show the LDOS on the edge a semi-infinite film for $h(\mathbf{k})$.
The finite size gap agrees with that obtained with 3D lattice in Fig.4 well.
When $L>20\mathring{A}{}$ the edge states are found connecting the valence
and conduction band. After the system becomes trivial when $L<20\mathring{A}{
}$, the edge states do not cross the band gap anymore, instead they only
attach to the conduction band. At the critical point $L=20\mathring{A}{}$, the
valence band and conduction band touch and form a linear Dirac cone at the low energy
regime, which is showed in both Fig.4 and Fig.5. This shows that by controlling the film thickness, it is possible to obtain a single valley Dirac cone for each spin block without using the topological surface states.\cite{Molenkamp-dirac}
Notice that in Fig.5 we can also see the edge states submerging in the valence bands, which
is similar to the case of second type surface states discussed previously. We call them the second type edge states.

\begin{figure}[t]
\centering \includegraphics[width=90mm]{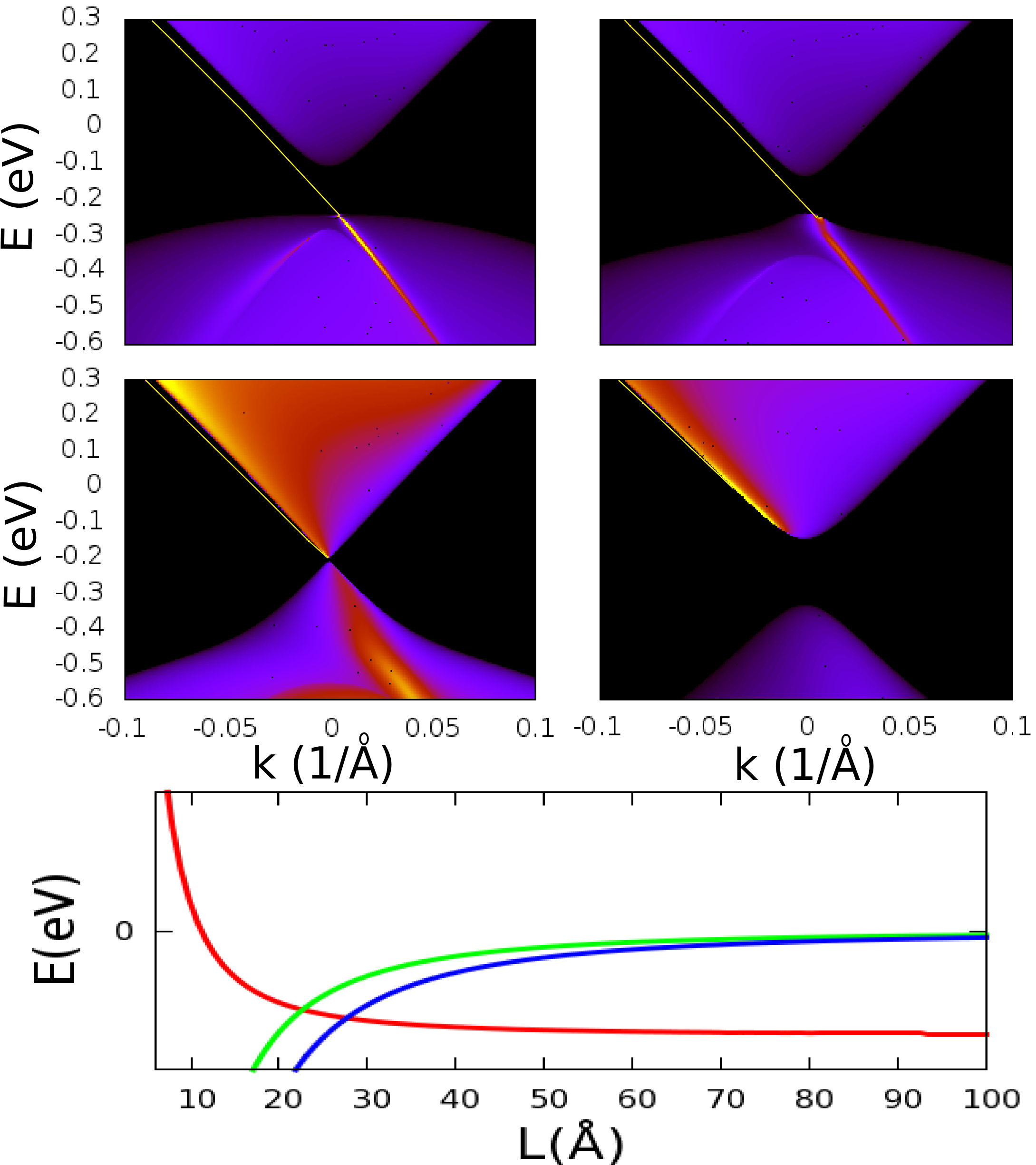}
\caption{(color online) LDOS at the edge of thin films at different thickness using
Hamiltonian (\protect\ref{upperHamiltonian}) with 2D lattice model. (a) L=28 
\AA {}; (b) L=25 \AA {}; (c) L=20 \AA {}; (d) L=16 \AA {}; (e) illustration of position change of LH, HH and $\Gamma_6$ bands as film thickness is varied.}
\end{figure}

\subsection{Chern number description of thin film band topology}

In Quantum Hall Effect(QHE), bulk-edge correspondence tells us that nonzero
Thouless-Kohmoto-Nightingale-den Nijs (TKNN) integer, summed by the Chern
number of occupied bands, is closely related to the presence of edge states
on the sample boundaries.\cite{Hatsugai-prl,Thouless-prl} 
For the time-reversal invariant systems which belong to the universality class of zero charge Chern number,
Z$_2$ index is then introduced to characterize the topologically nontrivial states.\cite{Kane, Qi-prb}
However, when the Hamiltonian of the system is composed of time-reversal diagonal blocks,\cite{Bernevig}
Z$_2$ invariant can be identified with the parity of the Chern number for each block.\cite{Fuprb}
Focusing on the
upper block of the thin film Hamiltonian Eq(\ref{2Dmodel})(i.e $h(\mathbf{k})$), we can discuss
its QHE. To identify the existence of edge states, we can calculate the
first Chern number of each band and sum up for all occupied bands. The Berry
curvature for each band is defined as\cite{MCChang,Xiao-rmp} 
\begin{equation*}
\Omega _{n}(k)=i\left( \left\langle \frac{\partial u_{n,k}}{\partial k_{x}}
\right\vert \left. \frac{\partial u_{n,k}}{\partial k_{y}}\right\rangle
-\left\langle \frac{\partial u_{n,k}}{\partial k_{y}}\right\vert \left. 
\frac{\partial u_{n,k}}{\partial k_{x}}\right\rangle \right) 
\end{equation*}
Then the Chern number is computed by integrating Berry curvature over the
first BZ: $C_{n}=\frac{1}{2\pi }\int_{BZ}d^{2}k\Omega _{n}(k)$. For each
block of the 2D lattice model, Chern number can be given to each of the
three bands. Tuning the thickness of the quantum well $L$ can inverse the
band structure near the $\Gamma $ point, resulting in the change of Chern
number for the touching bands. Several critical thickness for L can be found
when there's band touching at the $\Gamma $ point: $L\approx 30\mathring{A}$
and $L\approx 20\mathring{A}$. Numerical result of this integral divides the
parameter space into three regions: for $L\in (30\mathring{A},+\infty )$,
Chern number $C=1,0,-1$ arranged in sequence from the lowest energy band to
the highest(the same for the latter two cases); for $L\in (20\mathring{A},30
\mathring{A})$, Chern number $C=2,-1,-1$ and for $L\in (0,20\mathring{A})$,
Chern number $C=2,-2,0$. When Fermi level lies between the highest and the
middle energy band, the TKNN integer $N=1$ for $L\in (20\mathring{A},+\infty
)$ and $N=0$ for $L\in (0,20\mathring{A})$. The former case is topologically
nontrivial( referred to in Fig. 5a and 5b) and the latter is trivial
(referred to in Fig 5d)

\begin{figure}[t]
\centering \includegraphics[width=90mm]{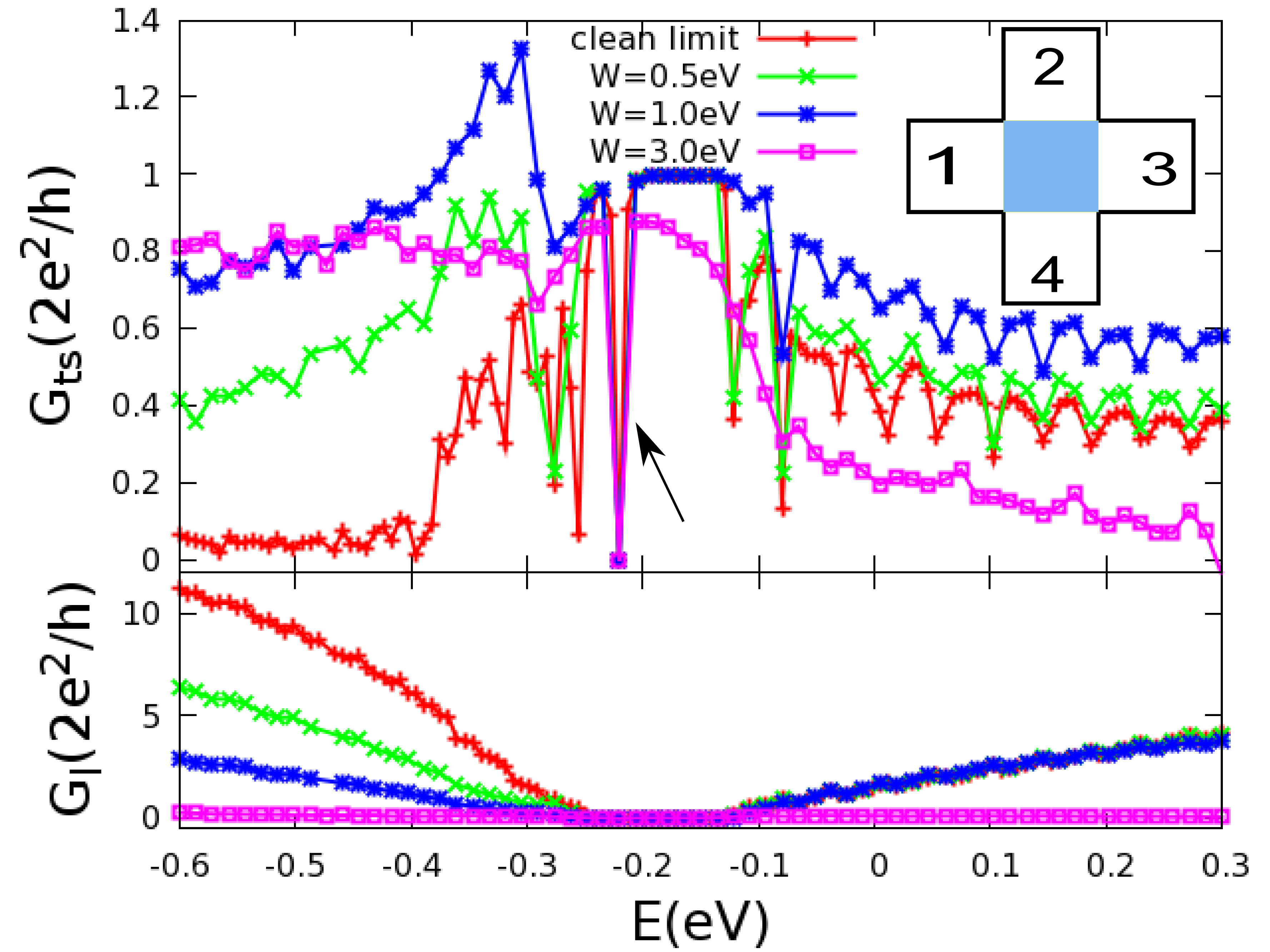}
\caption{(color online) Four terminal Landauer-B$\ddot{u}$ttiker calculations for thin
films. Four identical semi-infinite leads are attached to the cross bar
device, which is shown schematically in the insert. Disorders are put on the
central area(shaded areas). 200 disorder samples are taken for each disorder
strength $W$. The arrow indicates the position of finite size gap of the helical edge states.}
\end{figure}

\section{QSHE and Disorder enhanced spin Hall effect}

The edge state picture in thin films can be verified with explicit 
Landauer-B$\ddot{u}$ttiker calculations.\cite{Buttiker,Datta} We construct a
four-terminal Hall bar device using the tight binding version of Hamiltonian(\ref{2Dmodel}). 
Because the system keeps time reversal symmetry(TRS), we
need only consider a single block of the $6\times 6$ Hamiltonian. The transverse charge
conductance of each sub block is defined as $G_{tc}=G_{12}-G_{14}$.
$G_{ij}$ denotes the conductance between terminal $i$ and $j$, 
which is given by $G_{ij}=\frac{e^2}{h}Tr[\Gamma_iG^R\Gamma_jG^A]$. $\Gamma_i=i[\sum_i^R-\sum_i^A]$ is 
the spectrum function of lead i. $G^{R/A}$ is the retarded and advanced Green's function of the sample which 
has taken into account the four semi-infinite leads through the self energy $\sum_i^{R/A}$: $ G^{R/A}=[E-H_c-\sum_{i=1}^4\sum_i^{R/A}]^{-1}$. $H_c$ denotes the Hamiltonian of the shaded region of the sample.\cite{Datta,Hankiewicz,Li-05prb}
In real system with TRS, the transverse charge
conductance calculated here corresponds to the transverse 'spin' conductance, which is defined as $
G_{ts}=G_{tc}^{\uparrow }-G_{tc}^{\downarrow }$. The longitudinal conductance is defined as $G_l=G_{13}^{\uparrow}+G_{13}^{\downarrow}$. However, it should be noted that the 'spin' up or down here is only an indice of each Hamiltonian sub block and does not mean real spin. Because the basises of each sub block are hybrid of $j=\pm 1/2$ and $j=\pm 3/2$ states of $\Gamma_6$ and $\Gamma_8$ bands.\cite{Bernevig}

In Fig.6, we show the result
for a film whose thickness is 25 $\mathring{A}{}$. In the clean limit, i.e. without disorder, in the finite size
induced band gap the $G_l$ vanishes, $G_{ts}$ is quantized as $G_{ts}=2e^{2}/h$ (with $G_{12}^{\uparrow }=G_{14}^{\downarrow}=e^2/h$, $G_{14}^{\uparrow }=G_{12}^{\downarrow}=0$) , which proves the existence of helical edge states in the thin films and indicates the existence of quantum spin Hall effect(QSHE). \cite{Murakami-Science,Sinova-PRL,Shen-04prb,Sinova-05ssc,Kane-she}. In the
conduction band, the edge states coexist with the bulk states. $G_{ts}$
there is not quantized, which implies that the bulk states themselves carry
transverse spin conductance that partly cancels out the quantized spin Hall
conductance carried by the edge state channels. In the valence band, similarly,
the second type edge states coexists with bulk states. Notice that the
confinement in x,y dimension also generates a finite size gap in the edge
state itself, which is indicated by the arrow in Fig.6 where $G_{ts}$ vanishes. 

Disorder is recently known to play an important role in some exotic
phenomena in TIs.\cite{TAI,Beenaker} In real materials, dirty impurities are
inevitable. It is even possible to control them artificially.\cite
{Xue-PRL266102,Xue-PRL266803} Here we consider the effect of disorder
impurities to the transport properties of the thin films. Anderson type
white noise is introduced on the lattice model, which are spin independent on
site random potentials in the range $[-W/2,W/2]$. TRS is not voilated by disorder. 
In Fig.6 we show the transverse spin conductance $G_{ts}$ of the four terminal device at various
disorder strength $W$. The quantized $G_{ts}$ in the band gap shows
robustness against modest disorder strength, which is expected because the disorder we apply does not couple the TRS sub-blocks of the Hamiltonian and thus brings no backscattering between different spin edge channels.\cite{Kane,Bernevig}
Interestingly, $G_{ts}$ is significantly enhanced in both conduction and valence band even with weak
disorder. While $G_{ts}$ quickly becomes suppressed in the conduction
band when disorder increases, it still remains enhanced in the valence band.
Notice that with modest disorder strength(3eV in Fig.6), the longitudinal conductance $G_l$ already vanishes in the low energy regime, but the transverse spin conductance remains enhanced in the valence band.
Therefore with modest disorder strength, we expect a strong disorder enhanced spin Hall effect
in the valence band, which would manifest itself through strong signals in non-local measurements. In a very recent experiment similar observation of SHE difference in valence and conduction band has been reported with HgTe quantum wells.\cite{Molenkamp-natphys}

\section{Summary}

We have demonstrated that 6-band Kane model can be utilized to study the
topological properties of both 3D and thin film realistic materials. By choosing
proper model parameters, it describes the band structure well in the low
energy regime and captures the essential band topological features
sufficiently. By calculating the LDOS on the system boundary, the existence
of topological surface states is explicitly demonstrated. Using the model
parameters of HgTe, we show that even though the system is semi-metal 
surface states already exist in the material as long as the band structure
is inverted. We also find in the strain free condition surface states are
divided into two parts in the spectrum, each of different characteristics.
We demonstrate that uniaxial strains can generate an insulating gap and
transform the semi-metals into true TIs, in which the gapless Dirac cone of
surface states is found in the bulk band gap. Because of the
similar band structures and band topologies, we expect that the same physics
applies to the recently discovered Heusler-related and Li-based
intermetallic ternary compounds, which are mainly topologically nontrivial
semi-metals or metals.

We also demonstrate a crossover from 3D topological semi-metal to 2D QSHE insulators. 
In the thin films made of these materials, we predict the existence of
helical edge states and QSHE in the strain free condition. And finally we show that disorder plays an
important role in the transport properties in the thin films. It
significantly enhances the SHE in the valence bands. 

\section{Acknowledgment}

This work was supported by the Research Grant Council of Hong Kong under
Grant No.: HKU 7037/08P, HKUST3/CRF/09 and in part by a Hong Kong UGC
Special Equipment Grant (SEG HKU09).

\end{document}